\documentclass[twocolumn,twoside,slac_two]{revtex4}
\usepackage{graphicx}
\usepackage{fancyhdr}
\begin{document}

\title{The spectrum of charmed mesons from dynamical anisotropic lattices}

\author{A.~\'O Cais}
\affiliation{Centre for the Subatomic Structure of Matter, School of 
Chemistry \& Physics,\\
The University of Adelaide, Adelaide, SA 5005, Australia
}
\author{M. B. Oktay\footnote{Presenting Author}, S.~M.~Ryan, M.~J.~Peardon}
\affiliation{School of Mathematics, Trinity College, Dublin 2, Ireland}
\author{J.~I.~Skullerud}
\affiliation{ Department of Mathematical Physics, NUI Maynooth, Co. Kildare, Ireland}

\begin{abstract}
We present our preliminary analysis for the chamonium and D$_s$ spectra
obtained from N$_f=2$ dynamical anisotropic lattices. We use 12$^3\times 80$
lattices with lattice spacing $a_t=7.35$ GeV$^{-1}$ and anisotropy of six. 
Meson correlators are computed using all-to-all propagators together 
with variational analysis. 
\end{abstract}

\maketitle

\thispagestyle{fancy}

\section{INTRODUCTION}
In recent years, there has been renewed interest in charm physics. 
Many new states such as the X(3872), the Y(4260) and the $D_{sJ}$ have been observed 
\cite{Choi:2003ue, Aubert:2005rm, Abe:2004zs,Aubert:2003fg,Besson:2003jp}
and their precise measurement including $J^{PC}$ numbers, has become an 
important topic both experimentally and theoretically.

In principle, lattice QCD should be able to answer these questions from
first principles but it requires high precision numerical simulations.
In this region of the quark mass relativistic effects could be important.
However, simulations of the  charm quark with isotropic 
lattices are expensive. In this study we use a relativistic 
anisotropic lattice formulation with $N_f=2$ dynamical quarks to study 
charmonium and $D_s$ spectra. In this formulation, the lattice is discretized
along the spatial, $a_s$, and temporal, $a_t$, directions with $\xi=a_s/a_t\gg 1$.
Anisotropic lattices have the advantage of having small discretization
errors in the temporal direction whilst keeping the computational cost down. 
In addition, keeping 
a small temporal lattice spacing allows us to increase the number of
time slices which in turn makes it easier to identify the plateau
regions in effective mass plots. It is difficult to achieve 
this using isotropic lattices since the heavy hadron correlators with a charm 
or bottom quark have a signal that decays rapidly. Our aim is to be able
to extract the excited spectra with small statistical errors. 

Our lattice actions are described in section~\ref{sec:action}. 
In section~\ref{sec:simulation} we summarize the parameters and the
interpolating operators used in the simulation. In section~\ref{sec:analysis} 
we discuss the method to obtain excited spectra. In conclusion, we
discuss our results and future plans.

\section{ANISOTROPIC ACTIONS \label{sec:action}}
In this section, we describe the gauge and quark actions 
used in the simulation. The gauge action is a Two-Plaquette Symanzik-improved 
action which is designed to study glueballs \cite{Morningstar:1999dh}.
It is given by
\begin{eqnarray}
S_G&=&\frac{\beta}{\xi_g^0}\left\{
      \frac{5(1+w)}{3u_s^4}\Omega_s-\frac{5w}{3u_S^8}\Omega_{s}^{(2t)}
     -\frac{1}{12u_s^6}\Omega_s^{R}
\right\} \nonumber \\
   &+&\beta\xi_g^{0}\left\{\frac{4}{3u_s^2u_t^2}\Omega_t-\frac{1}{12u_s^4u_t^2}
    \Omega_t^{(R)} 
\right\} ,
\end{eqnarray} 
where $\Omega_{s} $ and $\Omega_t$ are the  spatial and temporal plaquettes
respectively. $\Omega_s^{(R)}$ and $\Omega_t^{(R)}$ refer to space-space and
space-time rectangles and 
$\Omega_s^{(2t)}=\frac{1}{2}\sum_{x,i>j}\left[1-P_{ij}(x)P_{ij}(x+\hat{t})
\right]$. This action has leading discretization errors of
 $O(a_s^4,a_t^2,\alpha_sa_s^2)$. 

The $O(a_s^3,a_t,\alpha_sa_s)$-improved anisotropic quark action 
\cite{Foley:2004jf} is given by
\begin{eqnarray}
S_q&=&\bar{\psi}\left[
      \gamma_0\nabla_0+\sum_i\mu_r\gamma_i\nabla_i\left(1-
      \frac{1}{\xi_q^0a_s^2}\Delta_i\right)\right. \nonumber \\
&-& \left. \frac{ra_t}{2}\Delta_{i0}
      +sa_s^3\sum_i\Delta_i^2+m_0
\right]\psi ,
\end{eqnarray}
where $r$ and $s$ are the Wilson parameters and $\mu_r=(1+ra_tm_0/2)$. The
derivatives are defined as
\begin{equation}
\nabla_\mu\psi(x)=\frac{1}{2a_\mu} \left[U_\mu(x)\psi(x\!+\!\hat{\mu})-
U_\mu^\dagger(x\!-\!\hat{\mu})\psi(x\!-\!\hat{\mu})\right] ,
\end{equation}
\begin{eqnarray}
\Delta_\mu\psi(x)&=&\frac{1}{a_\mu^2}\left[U_\mu(x)\psi(x\!+\!\hat{\mu})\!+\!
U_\mu^\dagger(x-\hat{\mu})\psi(x\!-\!\hat{\mu})\right. \nonumber \\
 & &-\!\left. 2\psi(x)\right] ,
\end{eqnarray}
and $\xi_q^0$ is the bare quark anisotropy. In order to maximize the
plaquette stout links \cite{Morningstar:2003gk} are used.
We used the same quark action to simulate light sea quarks and 
heavy valence quarks. Our sea quark mass in this simulation is close
to the strange quark mass. 

The ratio of the lattice spacings, $a_s/a_t$, which appears in both 
the gauge and the quark actions, as $\xi_g^{0}$ and $\xi_q^{0}$ respectively, 
are bare parameters that need to be tuned. In a
quenched study, this tuning can be done separately. However, sea quark 
effects in  dynamical simulations lead to a simultaneous 
non-trivial tuning of the anisotropies. This procedure is explained in
detail in Ref.~\cite{Morrin:2005tc}. For the results presented in these
proceeding the renormalized anisotropy, $\xi_r$ is set to be 6.

\section{SIMULATION DETAILS \label{sec:simulation}}
In this study, we obtained the charmonium and the $D_s$ spectra 
from $12^3\times 80$ lattices with 250 configurations. We tuned
the valence charm quark mass to $a_tm_c=0.117$ in order to get the
$J/\psi$ mass correct while the light quark mass is 
$a_tm_{sea}=a_tm_{light}=-0.057$
which is close to the strange quark mass. We use the all-to-all propagator 
method with ``dilution'' of Ref.~\cite{Foley:2005ac} with no eigenvalues
for the charm quark propagators and 20 eigenvalues for the strange quark 
propagators. We used time, space(even/odd) and color dilution 
for the charmonium study while color dilution is omitted in the $D_s$ case.
The lattice spacing is set from the spin-averaged (1P-1S) splitting 
in the charmonium system 
and found to be $a_t\simeq0.0272$ fm. The parameters used are listed in
table~\ref{tab:params}.

\begin{table}[t]
\begin{center}
\arraystretch{1.5}
\caption{Simulation parameters.}
\begin{tabular}{|l|l|}
\hline 
\textbf{Configurations} & 250 ($a_tm_c=0.117$, \\
                        & $a_tm_{sea}=a_tm_{light}=-0.057$)
\\
\textbf{Dilution}  & time, space(even/odd), color ($\bar{c}c$)\\
                   & time, space(even/odd) , ($D_s$)\\
\textbf{Physics}   & S, P and D waves, hybrids\\
\textbf{Volume}    & $12^3\times 80$ \\
\textbf{$N_f$}     & 2 \\
\textbf{$a_s$}     & $\sim0.17$ fm \\
\textbf{$a_t^{-1}$}& $7.35\pm0.03$ GeV \\
\hline 
\end{tabular} 
\label{tab:params}
\end{center}
\end{table}

Taking advantage of the all-to-all propagators, we use a variational
method \cite{Luscher:1990ck,Michael:1985ne} in order to get a better overlap 
with higher excited states where we use a spatially extended operator 
basis~\cite{Lacock:1996vy}. 
The lattice operators used in this study along with their putative continuum spin 
assignments are given in table~\ref{tab:ops}. We have assumed the simplest possible 
continuum assignment is correct. The validity of this assumption is under investigation. 
Using the operator basis in table~\ref{tab:ops}, for a given channel $R$ 
one can construct the matrix
\begin{equation}
C^{(R)}_{\alpha\beta}(t)=
\langle 0 | O^{(R)}_\alpha(t) O^{\dagger (R)}_\beta(0) | 0\rangle ,
\end{equation} 
where $\alpha,\beta = 1,\dots, n$ represent the different interpolating
operators, constructed by applying different levels of quark smearing. 
The different energy levels can then be obtained from 
\begin{eqnarray}
\lim_{t\rightarrow\infty}\lambda_\alpha(t,t_0)= e^{-(t-t_0)E_\alpha}
\left[ 1 + O(e^{-t\Delta E_\alpha})\right] ,
\end{eqnarray}
where $\lambda_\alpha$ are the eigenvalues of the matrix 
$C(t_0)^{-1/2}C(t)C(t_0)^{-1/2}$ and $t_0$ is some small reference time.
We performed single state fits to the diagonal elements in order
to extract the ground and excited states. 
\begin{table}[h]
\begin{center}
\caption{The operator basis used to obtain the $\bar{c}c$ and $D_s$ spectra. 
Definitions of the $s_i, p_i$ and $t_i$ are given in Ref.~\cite{Lacock:1996vy}.}
\begin{tabular}{||c|c|c|c||}
\hline
$J^{\rm PC}$ & ${}^{2S+1}L_J$ & STATE & OPERATORS \\
\hline
 $0^{-+}$ & ${}^1S_0$
                      & $\eta_c$,$\eta_c^{'}$
                      & $\gamma_5$,$\gamma_5\sum_i s_i$ \\
 $1^{--}$ & ${}^3S_1$
                      & J/$\psi$,$\psi$(2S)
                      & $\gamma_j$,$\gamma_j\sum_i s_i$ \\
\hline
\hline
 $1^{+-}$ & ${}^1P_1$
                      & $h_c$,$h_c^{'}$
                      & $\gamma_i\gamma_j$, $\gamma_5p_j$ \\
 $0^{++}$ & ${}^3P_0$
                      & $\chi_{c_0}$,$\chi^{'}_{c_0}$
                      & 1,$\vec{\gamma}\cdot\vec{p}$ \\
 $1^{++}$ & ${}^3P_1$
                      & $\chi_{c_1}$,$\chi^{'}_{c_1}$
                      & $\gamma_5\gamma_i$,$\vec{\gamma}\times\vec{p}$ \\
 $2^{++}$ & ${}^3P_2$
                      & $\chi_{c_2}$,$\chi^{'}_{c_2}$
                      & $\vec{\gamma}\times\vec{p}$,
                        $\gamma_1p_1\!-\!\gamma_2p_2$ \\
                      & & & $2\gamma_3p_3-\gamma_1p_1-\gamma_2p_2$ \\
\hline
\hline
 $2^{-+}$ & ${}^1D_2$
                      & $1{}^1D_2$
                      & $\gamma_5(s_1-s_2)$,$\gamma_5(2s_3-s_1-s_2)$\\
 $2^{--}$ & ${}^3D_2$
                      & $1{}^3D_2$
                      & $\gamma_j(s_i-s_k)$,$\gamma_1t_1-\gamma_2t_2$,\\
                      & & & $2\gamma_3t_3-\gamma_1t_1-\gamma_2t_2$ \\
 $3^{--}$ & ${}^3D_3$
                      & $1{}^3D_3$
                      & $\vec{\gamma}\cdot\vec{t}$ \\
\hline
\hline
 $1^{-+}$ & Hybrid & $q\bar{q}g$
                      & $\vec{\gamma}\times\vec{u}$\\
\hline
\hline
\end{tabular}
\label{tab:ops}
\end{center}
\end{table}

\section{ANALYSIS \label{sec:analysis}}
\begin{figure}[h]
\centering
\includegraphics[width=79.5mm,height=57mm,angle=0]{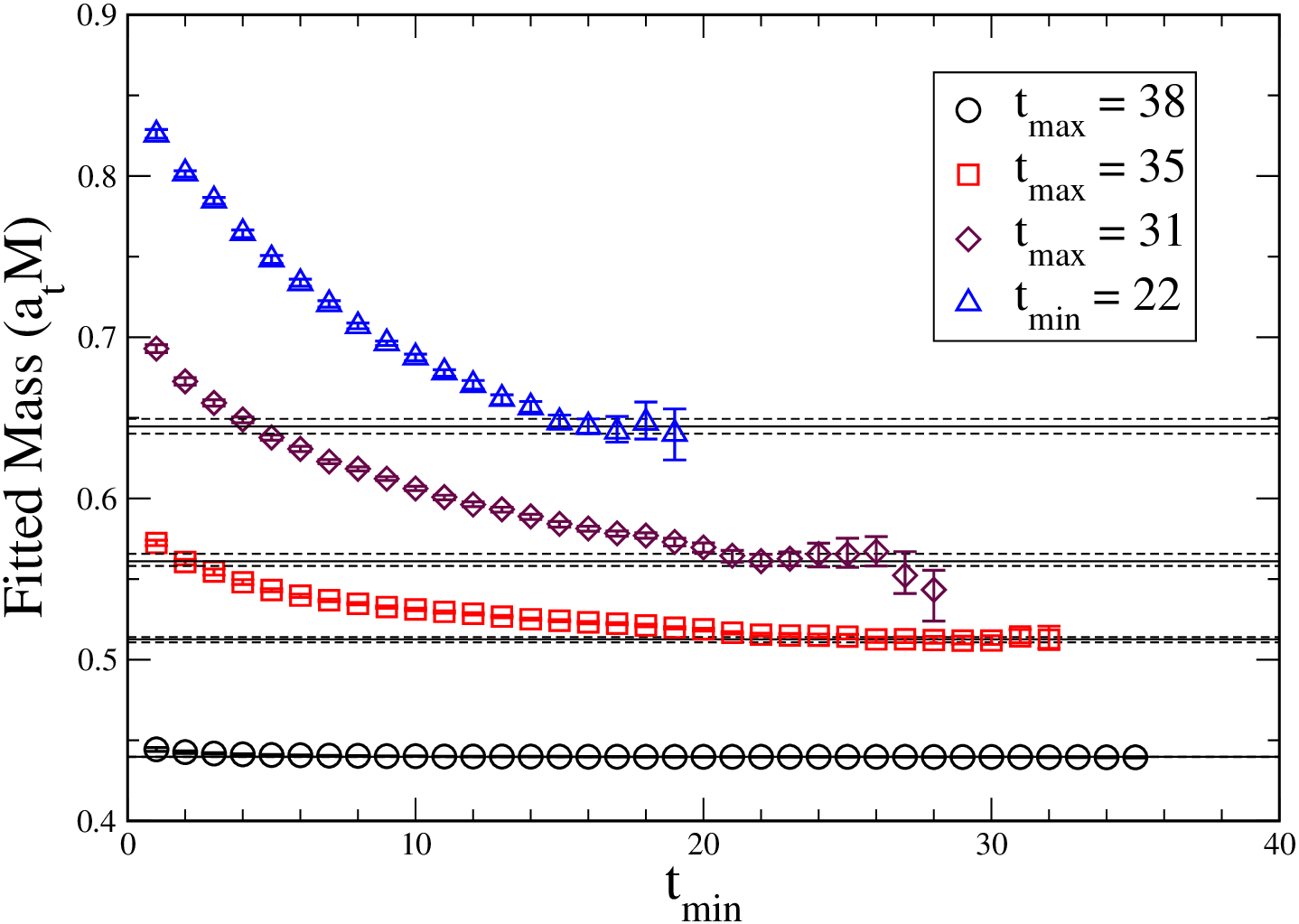}
\caption{Sliding window plots for the $J/\psi$ obtained from the variational
analysis. The plot shows the ground state in the lattice $T_{1_u}^-$ 
irreducible representation which corresponds to the continuum $J=1$ state. The 
higher lying states determined in this channel are assumed to be radial 
excitations of the $J/\Psi$. This is under investigation.}
\label{fig:jpsi}
\end{figure}

\begin{figure}[h]
\centering
\includegraphics[width=80.5mm,height=57mm,angle=0]{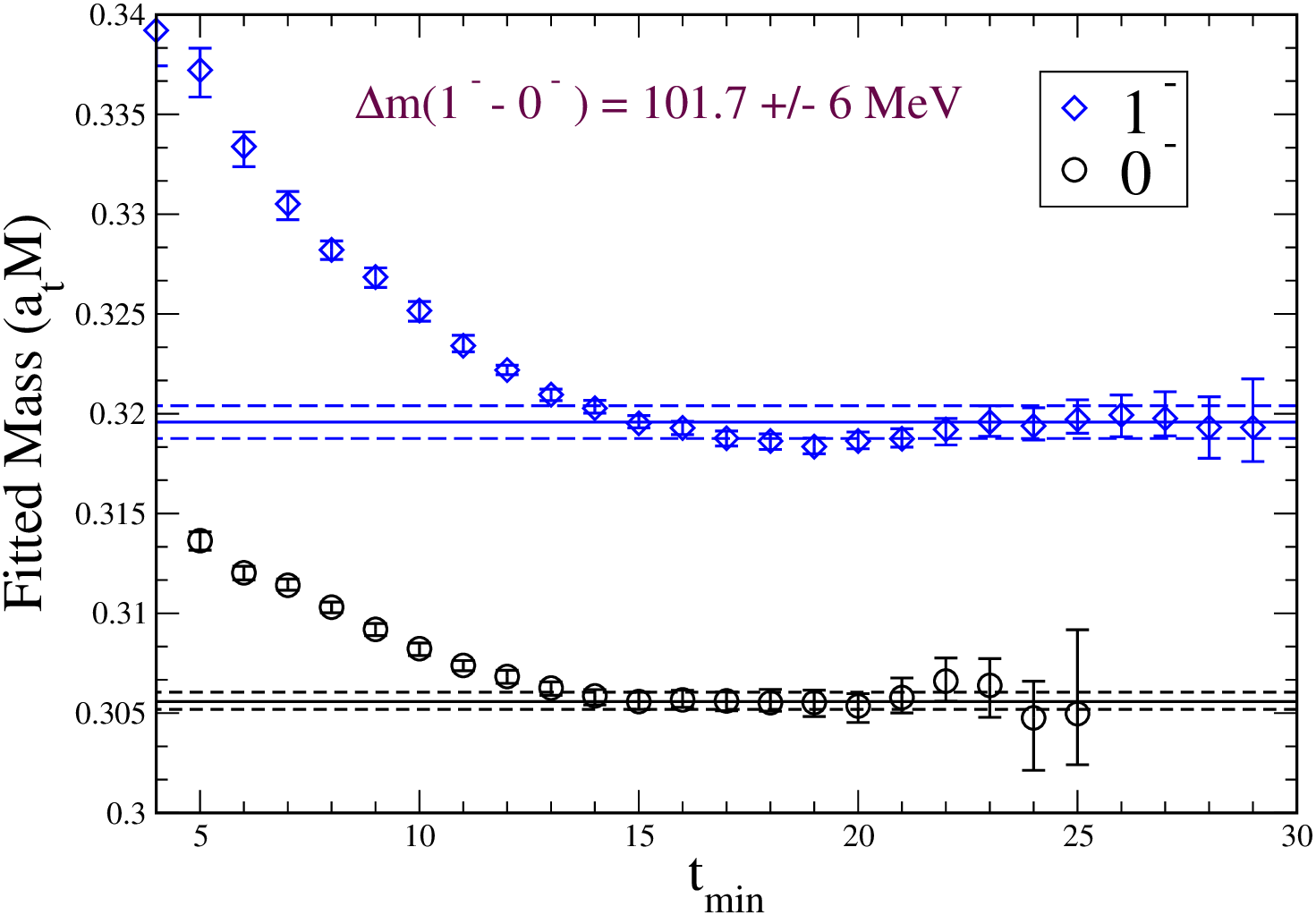}
\caption{Sliding window plot for the $D_s$ $0^-$ and $1^-$ states. The 
corresponding experimental value of the $1^--0^-$ splitting is 143.8MeV.}
\label{fig:Ds_slide}
\end{figure}

Time diluted all-to-all propagators introduce random noise at each
time slice which makes it difficult to indentify a plateau region
in the effective mass plots which fluctuate more than point-to-all 
propagators. However, we fit the correlators and a better picture can be 
obtained from ``sliding window'' plots 
(or $t_{min}$ plots). For a fixed value of some $t_{max}$, we
vary the $t_{min}$ value and plot the fitted mass. This is illustrated
in Figs.~\ref{fig:jpsi} and ~\ref{fig:Ds_slide} for the $J/\psi$ and $D_s$ $0^-$ and
$1^-$ states, respectively \cite{Juge:2005nr,Juge:2006fm}. 
We choose our fits based on the $\chi^2/N_{df}$ ($<$2), fit range
(where the fits are stable) and the fit quality (Q$>$0.2). These values
are chosen from our earlier simluations with smaller lattices. Most of
our results have better $\chi^2$ and $Q$ values. 

\begin{figure}[h]
\centering
\includegraphics[width=77.5mm,height=60mm]{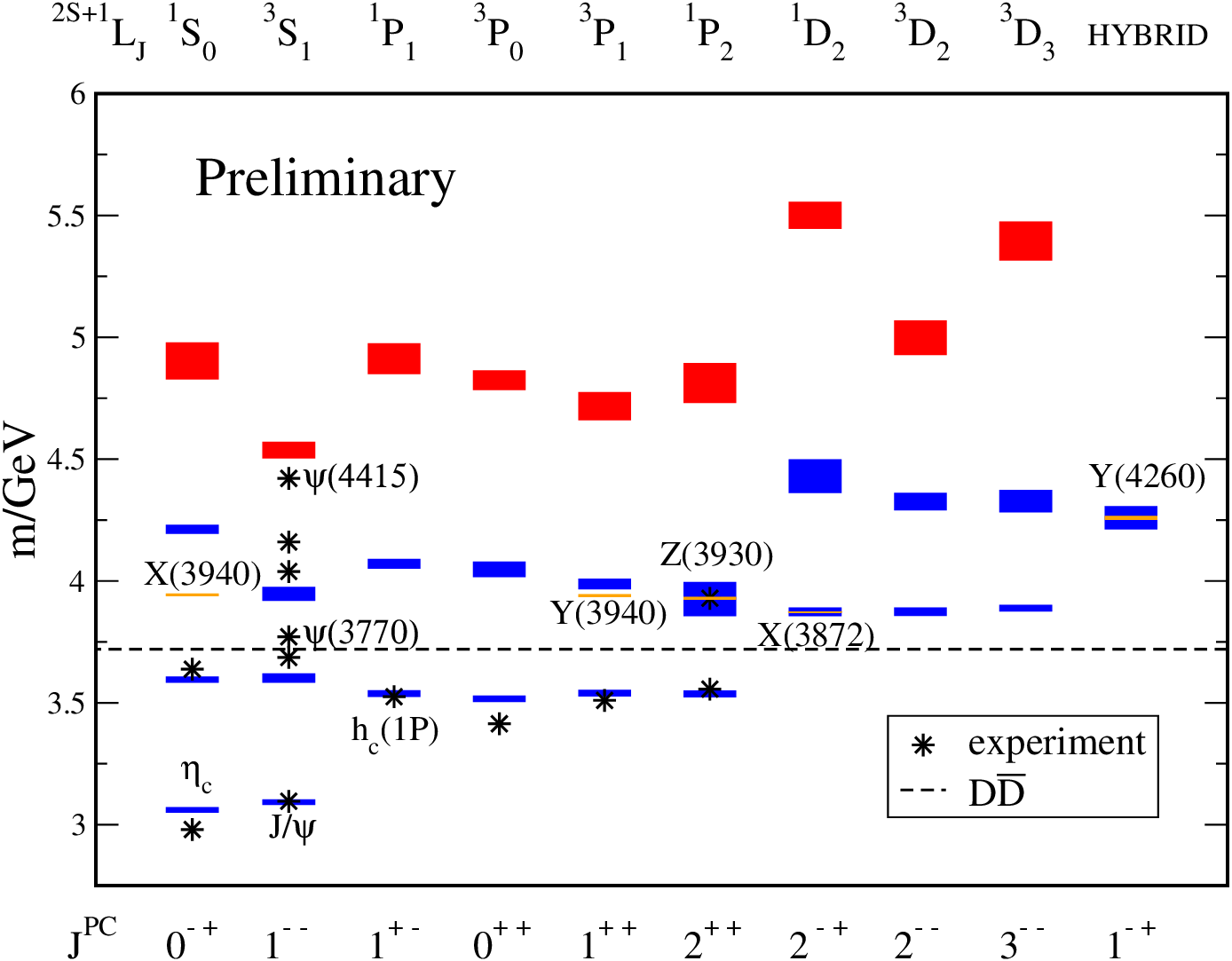}
\caption{Preliminary charmonium spectrum for the S, P, D waves and the
hybrid $1^{-+}$. The results of this study are the blue and red bands. The 
highest lying radial excitations identified in each channel are coloured red 
to indicate that these are not free of further excited-state contamination.} 
\label{fig:charmonium}
\end{figure}

\begin{figure}[h]
\centering
\includegraphics[width=82mm,height=57mm]{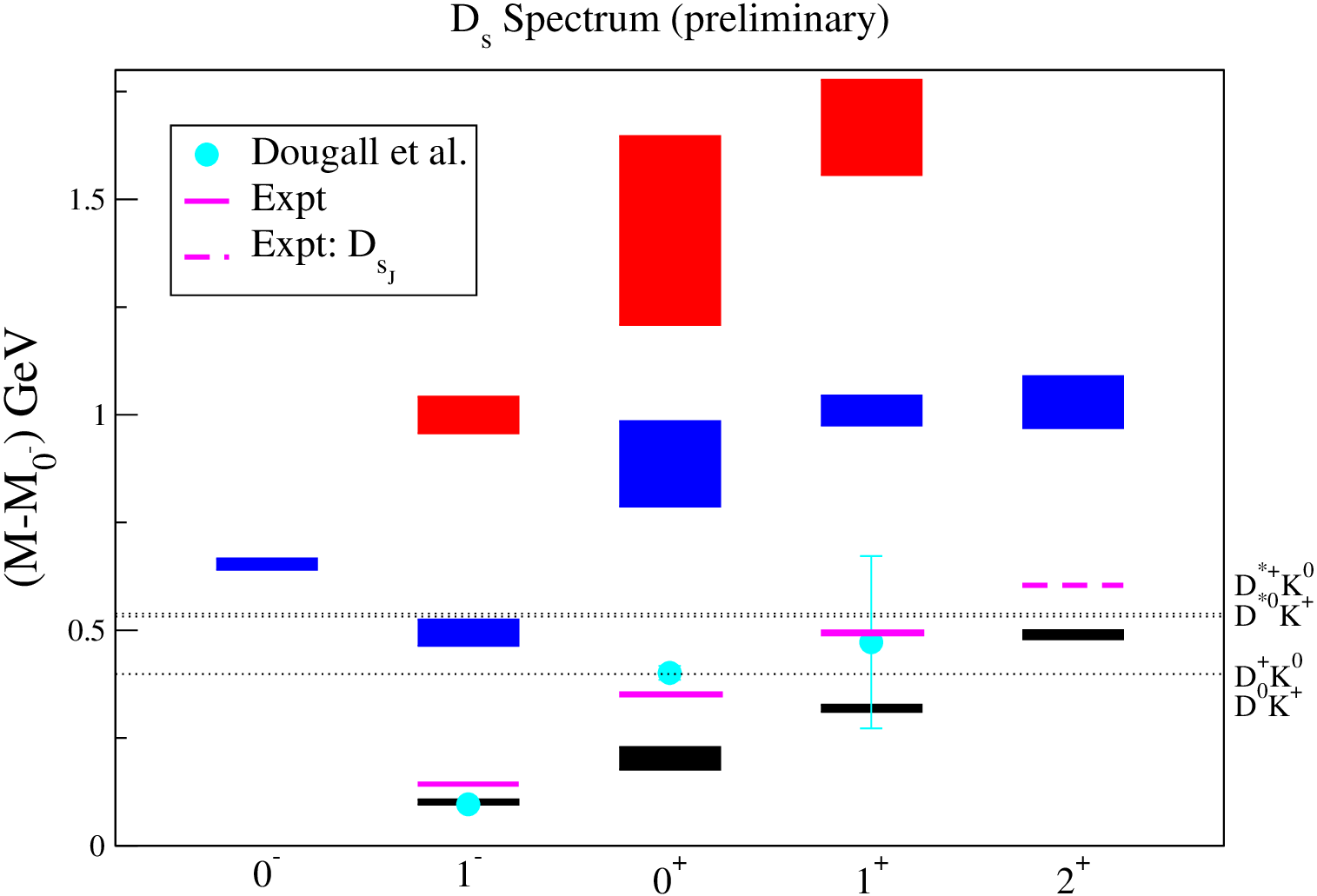}
\caption{Preliminary $D_s$ spectrum for the S and P waves. The results of this 
study are shown by the blue and red bands. As in the charmonium case the 
highest lying radial excitations identified in each channel are coloured red 
to indicate that these are not free of further excited-state contamination.  
The blue dots represent the UKQCD ($N_f=2$) results~\cite{Dougall:2003hv}.}
\label{fig:ds}
\end{figure}

We performed the same analysis for the $\bar{c}c$ and $D_s$ systems. Our
preliminary spectra are shown in figures~\ref{fig:charmonium}
 and~\ref{fig:ds}.

\section{Conclusions and Outlook}
We have presented our preliminary results for the charmonium and $D_s$
systems from $N_f=2$ dynamical anisotropic lattices. All-to-all
propagators are essential in this study and allow us to use a wide
range of operators and the variational analysis. For the charmonium
system we have good signals for the S, P and D waves and the $1^{-+}$
hybrid. We are planning to expand the simulation to include the $1^{--}$ 
D-waves
and other hybrids. We found the hyperfine splitting in this system 
to be small. The effect of the chromomagnetic term, 
$c_B{\mathbf\Sigma}\cdot{\mathbf B}$, 
disconnected diagrams and stout link smearing are being investigated as 
possible reasons for this. The $D_s$ system 
is simulated with a low level of dilution. A simulation with a 
higher level of dilution
for this system, with a wider range of operators, is being performed. 
Both simulations are 
performed at single lattice spacing where the sea quark mass is around
the strange quark mass. Simulations with finer lattices spacings are 
currently under investigation. The results clearly demonstrate the power 
of the all-to-all propagators combined with a variational analysis. We 
have extracted a 
large number of orbital and radial excitations in the charmonium and $D_s$
systems. Further work is also underway to address the continuum 
spin-identification of these lattice determinations.

\bigskip 
\begin{acknowledgments}
This work was supported by the IITAC project, funded by the Irish
Higher Education Authority under PRTLI cycle 3 of the National
Development Plan and by SFI grants 04/BRG/P0275, 04/BRG/P0266 and 
06/RFP/PHY061 and IRCSET grant SC/03/393Y.
\end{acknowledgments}

\end{document}